\documentclass[twocolumn,showpacs,preprintnumbers,amsmath,amssymb]{revtex4}

\usepackage{graphicx}
\usepackage{dcolumn}
\usepackage{bm}

\begin{document}

\title{Tackling master equations with a flux loop transform}

\author{S. Herminghaus}

\affiliation{Max-Planck-Institute for Dynamics and Self-Organization, Bunsenstr.
10, 37073 G\"ottingen, Germany}


\date{\today}

\begin{abstract}
A procedure is introduced which allows to represent the dynamics of
a non-equilibrium system violating detailed balance by its steady
state loop fluxes as the 'states'. Surprisingly, this loop
representation is found to obey detailed balance. A novel algorithm
for the construction of the steady state probability densities
naturally emerges, as well as a 'free energy' functional the
minimization of which yields the densities and currents of the
non-equilibrium steady state.
\end{abstract}


\maketitle

It is one of the major open questions of physics whether there
exists a general principle according to which systems far from
thermal equilibrium find their quasi-stationary states. While
equilibrium systems simply seek the minimum of the free energy, an
analogous functional governing non-equilibrium steady states (NESS)
is not known. The search for such functional has been launched in
the mid-twentieth century
\cite{Glansdorff71,Graham73,Keizer74,Haken75}, but in spite of the
many authors who have contributed to this demanding task (too many
to be mentioned here, in fact), it has so far not been successful.
The well-known major obstacle is the absence of detailed balance in
most systems of practical relevance, i.e., the presence of
non-trivial probability currents in the NESS \cite{Klein55}. The
present paper is not meant to add yet another trial to this list,
but to point out a subtle structure which is common to a wide class
of non-equilibrium systems, but appears to have been overlooked so
far. More specifically, a procedure is introduced which transforms a
system from a representation by its configurations into a
representation in the space of all possible closed loops
\cite{loopnote} of probability flux. Surprisingly, detailed balance
holds in this latter representation, providing a basis for defining
a statistical potential in a quite conventional manner. This may
turn out to be a useful link between equilibrium statistical physics
and the dynamics of systems far from thermal equilibrium.

Consider a system which can be in any one of $N$ configurations,
$\mathcal{C}_i$, with probabilities $p_i$. The system is fully
determined by the set of conditional probabilities $a_{ij}\tau
> 0$ to find the system in $\mathcal{C}_j$ at time $t+\tau$ provided
it was in $\mathcal{C}_i$ at time $t$, with a suitable (small) time
step $\tau$. We then have $\sum_j a_{ij} = 1 \ \forall i$. The
probability fluxes from state $i$ to state $j$ are
\begin{equation}
\phi_{ij}(t) = a_{ij} p_i(t)
\label{ProbabilityCurrents}
\end{equation}
which implies that there is no memory of past transitions, i.e., the
dynamics is considered a first-order Markov process. For most
processes of physical interest, this can be achieved by proper
definition of the states $\mathcal{C}_i$. The temporal evolution of
the probabilities is described by a master equation,
\begin{equation}
\frac{d p_i}{d t} = \sum_j ( p_j a_{ji} - p_i a_{ij}) \label{Master}
\end{equation}
The stationary solution of eq.~(\ref{Master}) defines the set of
steady state probabilities, $P^{\ast} := \{p_i^{\ast}\}$ (the
asterisk indicates steady state quantities throughout this paper).

The description outlined above is widely used for a huge range of
systems, including reaction-diffusion processes \cite{Baras96,Wu07},
systems biology \cite{Bhalla99,Lee04,Gillespie06,Hegland07}, cell
migration \cite{Cai06}, trading market dynamics
\cite{Scalas04,Chatterjee05}, and many other classical
non-equilibrium systems of general importance \cite{Haken75}. For
continuum systems, the Master equation is replaced by the
Fokker-Planck equation, which can be obtained from
eq.~(\ref{Master}) by means of the Kramers-Moyal expansion. However,
since the basic ideas of the present paper can be more clearly
outlined using the discrete model, we will leave continuum
descriptions to future work.

In a system with detailed balance, all currents, $j_{ij} :=
\phi_{ij}-\phi_{ji}$, vanish in the steady state, such that
\begin{equation}
p^{\ast}_i a_{ij} = p^{\ast}_j a_{ji} \ \ \forall i,j
\label{DetailedBalance}
\end{equation}
In this case, eq.~(\ref{DetailedBalance}) tells us that if one
$p^{\ast}_i$ is known, the neighboring $p^{\ast}_j$ can be
immediately obtained by multiplying $p^{\ast}_i$ with the ratio of
the corresponding transition rates, $a_{ij}/a_{ji}$. Continuing this
procedure through the whole system, we can determine each of the
probabilities, independently of the path we chose in doing so. This
may be viewed as an integrability condition \cite{Lebowitz99}.

The steady state density can then be derived from the potential
\cite{Zia07}
\begin{equation}
\mathcal{U}_i := -\ln(\Pi_{0i}/\Pi_{i0}) \label{PotentialDB}
\end{equation}
where $\Pi_{0i}$ is defined as the product of all rates $a_{ij}$
traversed in going step by step from a reference state,
$\mathcal{C}_0$, to the state of consideration, $\mathcal{C}_i$. We
then have the simple relation $p^{\ast}_i = \mathcal{N} \exp
(-\mathcal{U}_i)$ for all $i$, where $\mathcal{N}$ is to be
determined from the normalization condition,
\begin{equation}
\sum_i p_i = 1. \label{Normalizep}
\end{equation}
However, in most systems of interest eq.~(\ref{DetailedBalance}) is
not fulfilled. As a consequence, $\mathcal{U}_i$ as determined via
eq.~(\ref{PotentialDB}) would depend upon the path chosen for its
evaluation. In other words, a potential in the above sense does not
exist.


For the sake of clarity, we adopt a graph theoretical
representation. Consider a graph $G = (V,E) $, with vertices $v_i
\in V$, each of which corresponds to a configuration of the system,
$\mathcal{C}_i$. $E$ is the set of directed edges, $e_{ij}$,
connecting the vertices $v_i$ and $v_j$. Each edge is associated
with the corresponding rate, $a_{ij}$. Note that we intrinsically
consider a maximal graph, in which $e_{ij}$ exists for each pair
$(v_i,v_j)$, although the corresponding rate constant, $a_{ij}$, may
be zero for many edges. They may actually be thought of as
arbitrarily small but finite, in order to secure ergodicity, i.e., a
unique NESS. We thus have $|V| = N$, and $|E| = N^2$, since the
'tadpoles' $e_{ii}$ are included as well.

If eq~(\ref{DetailedBalance}) is not fulfilled, what we henceforth
assume, there will be finite currents in the steady state. We prove
that every {\it stationary} current, $J^{\ast} :=
\{j_{ij}^{\ast}\}$, can be represented as a superposition of flux
loops. By a flux loop $\mathcal{L}$ of length $s$ we mean a set of
$s$ vertices and $s$ directed edges which form a closed path which
is self-avoiding, i.e., no vertex is visited twice. This assures
that the number of possible loops, $M$, is finite provided $G$ is
finite. We find $M = \sum_{s=1}^N N!/s(N-s)!$ for the number of
distinct self-avoiding loops which can be formed in $G$, and $s_k$
is the number of steps of the loop $\mathcal{L}_k$. To each loop we
assign an intensity $m_k^{\ast}$, which means that the corresponding
loop contributes a flux of strength $m_k^{\ast}$ to {\it each} of
the edges $e_{ij} \in E_k$. We shall now prove that there exists at
least one set of numbers $\{ m_k^{\ast} \}$ such that the stationary
{\it flux} distribution, $\Phi^{\ast}$, is equal to the
superposition of loop fluxes with strengths $m_k$. This entails the
analogous (weaker) statement for the stationary current, $J^{\ast}$,
which is already known as the flow decomposition theorem
\cite{Ahuja93}.

We begin by singling out one vertex, say, $v_0$. The other $N-1$
vertices are to be envisaged on a circle around it, ordered
according to their index, such that all edges ending or starting at
$v_0$ are radial directed lines. Consider now the 'triloops'
$\mathcal{L}_i^3 := (\{
v_0,v_i,v_{i+1}\},\{e_{0i},e_{i(i+1)},e_{(i+1)0}\})$ of length
$s_i=3$. We first consider the loop $\mathcal{L}_1^3 = (\{
v_0,v_1,v_2\},\{e_{01},e_{12},e_{20}\})$ and set $m_1^{\ast} =
\phi_{01}^{\ast}$. This fully accounts for the flux
$\phi_{01}^{\ast}$, but also yields a contribution of
$\phi_{01}^{\ast}$ to the edge $e_{20}$. Next we consider the
'biloop' $\mathcal{L}_{2}^2 = (\{ v_0,v_2\},\{e_{02},e_{20}\})$ of
length $s_2=2$. Its strength shall be $n^{\ast}_2 =
\phi^{\ast}_{20}-\phi^{\ast}_{01}$, such that $\phi_{20}^{\ast} =
m^{\ast}_1 + n^{\ast}_2$ is fully accounted for as well. Setting now
$m^{\ast}_2  = \phi^{\ast}_{02} - n^{\ast}_2$ for the intensity of
$\mathcal{L}_2^3 = (\{ v_0,v_2,v_3\},\{e_{02},e_{23},e_{30}\})$, we
also account for $\phi^{\ast}_{02}$, and so on. We continue in this
way all around the circle, until we arrive at $m_{N-1}^{\ast}$. This
corresponds to the last triangular loop which is left. Since
$\Phi^{\ast}$ is by definition an equilibrium flux distribution, we
know that $\sum_i (\phi_{0i}^{\ast}-\phi_{i0}^{\ast})$ must vanish.
Furthermore, since the contribution of each loop into $v_0$ vanishes
as well, this balance is not affected by the loops $\mathcal{L}_i^2$
or $\mathcal{L}_i^3$. As a consequence, since $\phi_{01}^{\ast}$ is
already fully accounted for by $m_1^{\ast}$, $\phi^{\ast}_{10}$ {\it
must} be equal to $m_{N-1}^{\ast}$, and $n^{\ast}_1 = 0$. This shows
that all net fluxes to and from $v_0$ can be represented by a
superposition of the triloop and biloop fluxes containing $v_0$.

Now we remember all intensities $m_i^{\ast}$ and $n_i^{\ast}$ we
have determined so far, and subtract the corresponding fluxes from
$\Phi^{\ast}$, such that there are no fluxes to or from $v_0$ left.
All that remains are fluxes within the system $V \backslash v_0$.
When subtracting the fluxes represented by the loops containing
$v_0$, we never violated the flux balance at any vertex, since all
involved fluxes were loops, and therefore themselves balanced at
each vertex. As a consequence, the field of fluxes in the truncated
graph is again balanced, i.e., the sum of all fluxes to and from
each vertex vanishes. We can then disregard the vertex $v_0$, and
proceed considering only the remaining truncated graph with $N-1$
vertices. We single out one vertex again, and make all fluxes into
it vanish by subtracting triloop and biloop fluxes, as described
above. Note that we will not have to update any of the intensities
of the flux loops containing $v_0$, since $v_0$ (and thus all loops
containing it) do not anymore belong to the system under study.

This procedure can be repeated until a graph of just two nodes is
left, which is of course a single biloop. We thus have constructed a
set $m_k^{\ast}$ such that the superposition of the corresponding
loop fluxes is equal to $\Phi^{\ast}$. This proves that every
balanced flux field can be represented by a superposition of triloop
and biloop fluxes. {\it A fortiori}, it proves that $ \Phi^{\ast}$
can be represented by a superposition of a set of {\it any} loop
fluxes, without specifying their lengths, $s_k$. It should be noted
that we cannot assure that all loop intensities are positive.
However, in all systems of relevance, the number of distinct loops,
$M$, is much larger than the number of edges, $N^2$, such that the
choice of the $m_k^{\ast}$ representing a certain $\Phi^{\ast}$ is
far from unique. In many (if not in all) cases, it will be possible
to exploit this freedom to choose all $m_k^{\ast}$ non-negative
\cite{footnote3}. We finally note that in a system with detailed
balance, a trivial choice is to have only biloops,
$\mathcal{L}_i^2$, with strengths $m_i^{\ast} = p_i^{\ast}a_{ij} =
p_j^{\ast}a_{ji}$.


Next we make use of the graph representation to obtain a pictorial
idea of the dynamics of the system. In the ensemble picture, we may
imagine that the NESS consists of a very large number of actors
travelling on the graph $G$ step by step, each of which represents a
realization of the system. Between the steps, actors reside at the
vertices, and during each step (i.e., once each time interval
$\tau$) each actor on a vertex, $v_i$, traverses one of its outgoing
edges, $e_{ij}$, according to the value of the corresponding rate
constant, $a_{ij}$. Choosing the edge $e_{ii}$ is to stay at this
vertex for another time $\tau$. The fluxes, $\phi_{ij}$, just count
the total number of actors traversing the edge $e_{ij}$ in one time
step, not caring which realization they represent. Using the result
obtained above, we may thus represent the the steady state fluxes by
imagining that each actor is eternally orbiting a single loop,
$\mathcal{L}_k$, with the number of actors on each loop being
proportional to $m_k^{\ast}$. More precisely, in the NESS each loop
is occupied by $q_k^{\ast} := s_k m_k^{\ast}$ actors, having exactly
$m_k^{\ast}$ actors on each of its $s_k$ vertices.

It is clear that in reality, the realizations of the system (i.e.,
the actors) will choose random continuations at each step instead of
orbiting the loops. In other words, there is a random exchange of
actors between loops at each step. We may imagine each actor to
carry a ticket for the loop he is currently orbiting. After each
step, these tickets are exchanged randomly between actors at each
vertex, such that the actors are redistributed among the loops, and
thus among the outgoing edges of the respective vertex. If this
exchange is a microscopically balanced random process, what we will
henceforth assume, it fulfills detailed balance. This is the key
idea of the 'flux loop transform' to be presented.


Before we proceed, we define the functional $\chi(x,X)$, where $x$
is an object and $X$ is a set of objects, by
\begin{equation}
\chi(x,X) = \left\{ \begin{array}{l} 1 \ \  {\rm if} \ x \in X \\ 0
\ \ {\rm else}
\end{array} \right.
\label{Euler}
\end{equation}
This allows for convenient bookkeeping of which vertex belongs to
which loop. With the help of eq.~(\ref{Euler}), we can write
\begin{equation}
\phi_{ij}^{\ast} = \sum_k m_k^{\ast} \chi(e_{ij},E_k) \ \ \forall i,j
\label{Constraint1}
\end{equation}
and
\begin{equation} p_i^{\ast} = \sum_k m_k^{\ast} \chi(v_i,V_k) \
\ \forall i
\label{Constraint2}
\end{equation}
Together with eq.~(\ref{Normalizep}) this leads to the normalization
\begin{equation}
\sum_k s_k m_k^{\ast} = \sum_k q_k^{\ast} = 1 \ \ \
\label{Constraint3}
\end{equation}
By combining eqs.~(\ref{Constraint1}) and (\ref{Constraint2}), we
can express the transition rates, $a_{ij}$, by the loop intensities
as
\begin{equation}
a_{ij} = \frac{\sum_k m_k^{\ast} \chi(e_{ij},E_k)}{ \sum_k
m_k^{\ast} \chi(v_i,V_k)}
\label{TransitionsByLoops}
\end{equation}
For now, however, we are still faced with the inverse problem, which is to
determine the $M$ numbers $m_k^{\ast}$ from the just $N^2+N+1$
equations~(\ref{Constraint1}), (\ref{Constraint2}), and
(\ref{Constraint3}).

This ambiguity may be greatly reduced by demanding the
representation of the fluxes in the space of loops to be optimized
in some sense, e.g., such as to prefer few large loops over many
small ones. This can be achieved by introducing a 'penalty
function', $I(\{m_k^{\ast}\}) := \sum_k g(s_k) m_k^{\ast \gamma}$,
where $g(s) > 0$ and $\gamma$ can be chosen freely. This freedom of
choice reflects the fundamental impact of the observer in defining
convenient coarse-grained variables for characterizing a
'self-organized' state. $I(\{m_k^{\ast}\})$ will be larger if more
loops are used to represent a certain flux field. A $g(s)$ which is
strongly decreasing will favor long loops, and a large exponent
$\gamma$ prevents too intense loop fluxes to appear. As we will see,
the choice of $g(s)$ and $\gamma$ will have no effect on the
predictions we make on the NESS acquired by the system under study.
As a particularly convenient choice, we set $g(s)=1$ and $\gamma =
2$. One might as well employ a 1-norm, $I(\{m_k^{\ast}\}) := \sum_k
g(s_k) |m_k|$, which leads to particularly few non-zero loop
intensities \cite{Boyd04,Timme07}.

If we now require $I(\{m_k^{\ast}\})$ to be minimal under the {\it
constraints}~(\ref{Constraint1}), (\ref{Constraint2}), and
(\ref{Constraint3}), we directly obtain, by means of the Lagrange
method,
\begin{equation}
m_k^{\ast} + \sum_{ij} \lambda_{ij} \chi(e_{ij},E_k) +
\sum_i \mu_i \chi(v_i,V_k) + \nu s_k = 0 \ \ \forall k\\
\label{Lagrange}
\end{equation}
where $\lambda_{ij}$, $\mu_{i}$, and $\nu$ are Lagrange multipliers.
Combining eqs.~(\ref{Constraint1}) and (\ref{Constraint2}), we
obtain
\begin{equation}
\sum_k \left(\chi(e_{ij},E_k) - a_{ij}
\chi(v_i,V_k)\right)m_k^{\ast} = 0 \ \ \forall i,j
\label{Constraint4}
\end{equation}
Inserting eq.~(\ref{Lagrange}) into eq.~(\ref{Constraint4}) leads to
$N^2$ equations for the set of $N^2 + N + 1$ Lagrange multipliers.
The latter are thus still under-determined, although to a lesser
degree than were the $m_k^{\ast}$.  For our choice of $\gamma = 2$,
however, $I$ just represents the distance to the origin in
$m^{\ast}$-space. Furthermore, the above equations for the
$m_k^{\ast}$ are all linear and thus define a hyperplane.
Consequently, minimization of $I$ under the above constraints just
amounts to finding the point within a hyperplane which is closest to
the origin (which is unique). The freedom in the Lagrange
multipliers thus defines a manifold within which the set of steady
state intensities is invariant, and a solution for ${m_k^{\ast}}$ is
uniquely obtained from $\{a_{ij}\}$ by the procedure above.

Once $\{m_k^{\ast}\}$ is known, the quantities characterizing the
NESS, $P^{\ast}$ and the corresponding currents $J^{\ast}$
\cite{Zia07}, can be computed directly from eqs.~(\ref{Constraint2})
and (\ref{ProbabilityCurrents}). Since all equations required to
obtain $\{m_k^{\ast}\}$ have a particularly simple structure, and
only contain the rate constants, this may be seen as an alternative
procedure to find the steady state solution to any master equation,
once the $a_{ij}$ are given. Its practical merits as compared to
other techniques, such as inversion of the matrix of rate constants,
$(a_{ij})$, or the method of directed trees \cite{Zia07}, remain to
be explored. Furthermore, it should be investigated if
$I(\{m_k^{\ast}\})$ can be chosen such as to guarantee a set of
non-negative $m_k^{\ast}$. These questions will be addressed in a
forthcoming paper.

Let us finally turn to the flux loop transform. Consider a graph, $H
= (W,F)$, with vertices $w_k \in W$, each of which corresponds to a
self-avoiding loop, $\mathcal{L}_k$, in $G$. $F$ is the set of
directed edges, $f_{kl}$, connecting the vertices $w_k$ and $w_l$.
To each vertex we assign the occupation number $q_k$ of the
corresponding loop, and each edge is associated with a transfer rate
constant, $b_{kl}$. We shall call the operation $G \longrightarrow
H$ the {\it loop transform}. By virtue of
eq.~(\ref{TransitionsByLoops}), its inverse exists and is unique.

To investigate the properties of the transformed graph, $H$, we need to specify
the numbers $\{b_{kl}\}$. The probability for an actor to transfer from loop
$\mathcal{L}_k$ to loop $\mathcal{L}_l$, at a certain vertex $v_i$
(of $G$) which is common to $\mathcal{L}_k$ and $\mathcal{L}_l$
($v_i \in V_k \cap V_l$) can be directly written down considering
the ticket exchange process described above. It is
\begin{equation}
t_{kl}^{(i)} = \frac{m_l^{\ast}}{\sum_{k^{\prime}}
m_{k^{\prime}}^{\ast}
\chi(v_i,V_{k^{\prime}})}=\frac{m_l^{\ast}}{p_i^{\ast}}
\label{TransitionRatePartial}
\end{equation}
Since each vertex which is common to both loops yields an
independent chance to transfer from $\mathcal{L}_k$ to
$\mathcal{L}_l$, the total probability to do so is
\begin{equation}
t_{kl} = \sum_i t_{kl}^{(i)} \chi(v_i, V_k \cap V_l) = C_{kl}
m_l^{\ast}
\label{TransitionRateTotal}
\end{equation}
where $C_{kl} := \sum_i \chi(v_i, V_k \cap V_l)/p_i^{\ast}$. On the
other hand, since the probability for an actor on loop
$\mathcal{L}_k$ to be at a certain vertex $i$ is $1/s_k$, the rate
constant for transfer from $\mathcal{L}_k$ to $\mathcal{L}_l$ is
given by $b_{kl} = t_{kl}/s_k$. Since evidently $C_{kl}=C_{lk}$, it
is clear from eq.~(\ref{TransitionRateTotal}) that
\begin{equation}
q_k^{\ast} b_{kl} = q_l^{\ast} b_{lk} \label{DetailedBalanceH}
\end{equation}
which states that there is {\it detailed balance} in $H$. We can
thus apply eq.~(\ref{PotentialDB}), replacing $a_{ij}$ by $b_{kl}$,
to obtain a potential $\mathcal{H}_k$, such that the occupation numbers
$q_k^{\ast}$ are given by
\begin{equation}
q_k^{\ast} = \exp(-\mathcal{H}_k) \label{BoltzmannQ}
\end{equation}
Note that $\sum_k \exp (-\mathcal{H}_k) \equiv 1$ by normalization,
eq~(\ref{Constraint3}).

There is still a disfigurement in using $H$ to represent the system
under study. Since $H$ concerns only loop fluxes, all possible
densities $Q = \{ q_k \}$ represent flux fields $\Phi$ (in $G$)
which are balanced at each vertex, and thus yield a time-independent
density field, $P$. This does clearly not fulfill the master
equation~(\ref{Master}) in general, which shows that any dynamics in
$H$, away from the steady state, has nothing to do physically with
the dynamics in $G$ as described by eq.~(\ref{Master}). It is so far
only the equilibrium state of $H$ which has a physical meaning,
namely to represent the NESS in $G$.

This can be amended, however, by alleviating the requirement that
the $q_k$ actors on a loop $\mathcal{L}_k$ be evenly distributed
among it vertices, as is the case in the steady state. In what
follows, we thus distinguish between the occupation numbers
$q_{k,i}$ at the different vertices $v_i$ along the loop
$\mathcal{L}_k$. The relation $\sum_i q_{k,i} = s_k m_k = q_k$ is
fulfilled by definition. Furthermore, we slightly change the rule
for the random redistribution of the tickets introduced above such
that between any two time steps, all actors at each vertex give
their tickets away, and draw new ones with a probability according
to the corresponding equilibrium intensities,
$m_k^{\ast}/p_i^{\ast}$. In the steady state, this rule yields
identical results as the one before, but assures the correct
dynamics away from the NESS. Any distribution $P = \{p_i\}$ can then
be uniquely represented setting
\begin{equation}
q_{k,i} = \frac{p_i}{p_i^{\ast}}m_k^{\ast} \chi(v_i,V_k) \label{Qki}
\end{equation}
It should be noted that the temporal evolution of the $q_{k,i}(t)$
in $H$ does not simply obey a Master equation analogous to
eq.~(\ref{Master}). But the the dynamics of the system is, by means
of the flux loop transform, presented such that all violations of
detailed balance are concealed within the ('interior degrees of
freedom' of the) vertices of $H$. As time proceeds, not only will
the $m_k$ approach their equilibrium values, but also will the
occupation numbers of the different vertices corresponding to the
same loop approach each other ($q_{k,i}^{\ast} = m_k^{\ast} =
q_k^{\ast}/s_k$).

Remembering that the $q_k = \sum_i q_{k,i}$ approach the Boltzmann
distribution (\ref{BoltzmannQ}), we see that the system as
represented in $H$ is in some sense analogous to a classical system
having $M$ discrete levels at energies $\mathcal{W}_k :=
\mathcal{H}_k + \ln s_k$, each of which is $s_k$-fold degenerate.
The only difference is that in $H$, the $q_{k,i}$ perform a
'round-dance' within each energy level (i.e., in the vertices of
$H$), which becomes insignificant as the steady state is reached.
There is thus no physical manifestation of the violations of
detailed balance in the NESS as represented in $H$.

Finally, we may write down an entropy in $H$ as
\begin{equation}
\mathcal{S}(t) = - \sum_{k,i} q_{k,i} \ln q_{k,i} \label{Entropy}
\end{equation}
which depends on time via the $q_{k,i}$ if the system is started
away from the NESS. The latter will maximize $\mathcal{S}$ while
obeying (\ref{BoltzmannQ}). Defining thus $\mathcal{W} := \sum_k q_k
\mathcal{W}_k$, it is readily checked that minimizing
\begin{equation}
\mathcal{F} = \mathcal{W} - \mathcal{S}
= \sum_{k,i} q_{k,i}\ln\frac{q_{k,i}}{m_k^{\ast}} \label{FreeEnergy}
\end{equation}
which is very much reminiscent to a usual free energy functional,
yields precisely the NESS, i.e., the pair $\{P^{\ast},J^{\ast}\}$
\cite{Zia07}. It may thus be hoped that the loop transform outlined
above opens up the possibility to apply standard equilibrium
statistics formalism to a wide class of systems far from thermal
equilibrium.

The author is indebted to J\"urgen Vollmer, Vasily Zaburdaev,
Manfred Denker, Folkert M\"uller-Hoissen, Martin Brinkmann, Barbara
Drossel, and Axel Fingerle for inspiring discussions and numerous
helpful hints.


\begin{references}
\bibitem{loopnote} many authors use the term 'cycle' instead of
'loop'. I have decided for the latter term because its alludes to
something very simple (it is, actually, the 'atom' of the theory
presented here), in contrast to the latin/greek derived word
'cycle', which usually means complex things (life cycle etc.).
\end{references}

\end{document}